\documentclass[journal]{IEEEtran}
\usepackage[utf8]{inputenc}
\usepackage{color}
\usepackage{amsmath,graphicx,amssymb,mathtools}
\usepackage[hidelinks]{hyperref}
\DeclareMathOperator*{\argmin}{argmin}
\newcommand{\tab}{\hspace{.1in}}

\definecolor{colorgreen}{rgb}{0.05, 0.5, 0.06}

\newcommand{\vera}[1]{\textcolor{black}{#1}}
\newcommand{\veraa}[1]{\textcolor{black}{#1}}
\newcommand{\verab}[1]{\textcolor{black}{#1}}
\newcommand{\verac}[1]{\textcolor{black}{#1}}

\def\cpx{{\mathbb{C}}}
\def\real{{\mathbb{R}}}

\def\W{{\Gamma}}
\def\bo{\bm\omega}
\usepackage{verbatim,xspace,amssymb}
\usepackage{longtable}
\usepackage{tikz,pgfplots,array}
\pgfplotsset{compat=1.13}
\usetikzlibrary{shapes}
\usepackage{bm}
\usepackage{etoolbox}
\makeatletter
\patchcmd{\@makecaption}
  {\scshape}
  {}
  {}
  {}
\makeatletter
\patchcmd{\@makecaption}
  {\\}
  {.\ }
  {}
  {}
\makeatother

\begin{document}
\title{Efficient Regularized\\ Field Map Estimation in 3D MRI}

\author{Claire~Yilin~Lin,~\IEEEmembership{Student Member,~IEEE}
        and~Jeffrey~A.~Fessler,~\IEEEmembership{Fellow,~IEEE}
\thanks{This work was supported in part by NIH grants R01 EB023618, U01 EB026977, R21 AG061839, and NSF grant IIS 1838179.}
\thanks{C. Y. Lin is with the Department
of Mathematics, University of Michigan, Ann Arbor,
MI, 48109 USA (e-mail: yilinlin@umich.edu).}
\thanks{J. A. Fessler is with the Department of Electrical Engineering and Computer Science, University of Michigan, Ann Arbor, MI 48109 USA (e-mail: fessler@umich.edu).}
}

\markboth{IEEE TRANSACTIONS ON COMPUTATIONAL IMAGING}
{Shell \MakeLowercase{\textit{et al.}}: Efficient Regularized Field Map Estimation in 3D Multi-coil MRI}

\maketitle

\begin{abstract}
Magnetic field inhomogeneity estimation is important in some types of magnetic resonance imaging (MRI), including field-corrected reconstruction for fast MRI with long readout times, and chemical shift based water-fat imaging.
Regularized field map estimation methods that account for phase wrapping and noise involve nonconvex cost functions that require iterative algorithms.
Most existing minimization techniques were computationally or memory intensive for 3D datasets, and are designed for single-coil MRI.
This paper considers 3D MRI with \vera{optional consideration of} coil sensitivity, and addresses the \vera{multi-echo field map estimation and water-fat imaging problem.}
Our efficient algorithm uses a preconditioned nonlinear conjugate gradient method based on an incomplete Cholesky factorization of the Hessian of the cost function, along with a monotonic line search.
Numerical experiments show the computational advantage of the proposed algorithm over state-of-the-art methods with similar memory requirements.
\end{abstract}

\IEEEpeerreviewmaketitle

\section{Introduction}\label{sec:intro}
In magnetic resonance imaging (MRI), scans with long readout times require correction for magnetic field inhomogeneity during reconstruction to avoid artifacts~\cite{sekihara1985nmr}-\cite{funai2008regularized}.
Field inhomogeneity is also a nuisance parameter in chemical shift based water-fat imaging techniques~\cite{glover1991three}-\cite{bley2010fat}.
Field map estimation is thus crucial to field-corrected MR image reconstruction, and for fat and water image separation.

One \verac{field map estimation} approach is to acquire MR scans at multiple echo times \vera{(usually 2 or 3), where a small echo time difference can help resolve any phase wrapping issues and a large echo time difference
can help improve SNR.
One can then estimate field inhomogeneity using images reconstructed from these scans~\cite{funai2008regularized}.}
Since field maps tend to be smooth within tissue, estimation methods with smoothness assumptions have been proposed \verac{for water-fat separation}, including region growing techniques~\cite{reeder2005iterative}-\cite{doneva2010compressed}, filtering~\cite{windischberger2004robust}, curve fitting~\cite{schneider1991rapid}-\cite{sharma2013chemical}, multiresolution and subspace approaches~\cite{sharma2013chemical}-\cite{tsao2013hierarchical}, and graph cut algorithms~\cite{bioucas2007phase}.
\verac{To improve robustness of water and fat separation and reduce ambiguity of assignment, field map pre-estimation methods such as demodulation~\cite{diefenbach2018improving} and magnetization transfer~\cite{samsonov2019resolving} have been proposed as part of the water-fat imaging framework.}
Most of these methods, however, use various approximations to account for phase wrapping between different acquisitions.
In contrast, regularized estimation methods~\cite{funai2008regularized},\cite{hernando2008joint}-\cite{{hernando2010robust}} have been proposed to account for both phase wrapping and the smoothness of the field map from multiple acquisition images. 
Because the field map affects image phase, these approaches involve a nonconvex optimization problem that requires iterative methods. 

To solve such optimization problems,~\cite{funai2008regularized},\cite{huh2008water},\cite{allison2012accelerated} use a \vera{majorization}-minimization (MM) approach by introducing a quadratic majorizer for their cost functions.
The MM approach decreases the cost monotonically, but is computationally intensive, especially for large-scale datasets.
Other regularized field map estimation minimization techniques quantize the solution space~\cite{hernando2008joint},\cite{hernando2010robust} and may require a second descent algorithm to produce sufficiently smooth estimates. 
An alternative minimization technique~\cite{ongie2017efficient} uses nonlinear conjugate gradient (NCG) with a monotonic line search (MLS), and explored  various preconditioners in the 3D single-coil case. 

This paper considers the regularized field map estimation problem in the 3D \verac{multi-coil} MRI setting.
In particular, we consider a generalized cost function in the multi-coil case for both multi-echo field map estimation and water-fat imaging.
We minimize it by a NCG algorithm with an efficient MLS and an iteration-dependent preconditioner based on an incomplete Cholesky factorization~\cite{manteuffel1980incomplete} of the Hessian of the cost function.
\vera{The incomplete Cholesky factorization has been applied to field inhomogeneity estimation using surface fitting \cite{lai2003dual}, and recently to single-coil field map estimation with a similar cost function  \cite{ongie2017efficient}.}
In addition to faster convergence, this preconditioner exploits the sparse structure of the Hessian, thus it is memory efficient and scales to 3D datasets.
Compared to previous works~\cite{huh2008water},\cite{allison2012accelerated},\cite{ongie2017efficient}, our new approach unifies the field map correction and the water-fat imaging problems, with a generalized expression that \vera{optionally} considers multiple coils in MRI.
Our efficient algorithm on this problem shows significant computational and storage advantages
compared with existing MM and NCG methods.

The rest of this paper is organized as follows. 
Section~\ref{sec:prob} describes the optimization problem for the field map estimation problems for \verac{multi-coil} MRI.
Section~\ref{sec:alg} presents the NCG-MLS optimization scheme with the proposed preconditioner.
Section~\ref{sec:res} reports simulated and real experimental results, followed by conclusions in Section~\ref{sec:con}.

\section{Problem Formulation}\label{sec:prob}

We are given reconstructed images $\bm y_{cl} \in\cpx^{N_\mathrm{v}}$ for the $c$th receiver coil of the $l$th scan, with $c = 1,\ldots, N_\mathrm{c}\mskip2mu,$ $l = 1,\ldots,L$, where $N_\mathrm{v}$ denotes the total number of voxels in the image, $N_\mathrm{c}$ denotes the number of coils, and $L\geq 2$ denotes the number of echo times.
We model the field inhomogeneity effect as 
\begin{equation}
y_{clj}= e^{i\omega_j t_l}s_{cj}x_{lj} + \epsilon_{clj}\mskip2mu,
\label{eq:model}
\end{equation}
where $j=1,\ldots,N_\mathrm{v}$ is the voxel index, $\bo\in\real^{N_\mathrm{v}}$ is the unknown field map, $t_l\in\real$ is the echo time shift of the $l$th scan, $\bm s_c\in\cpx^{N_\mathrm{v}}$ is the (known) coil sensitivity map for the $c$th coil, and $\bm\epsilon_{cl}\in\cpx^{N_\mathrm{v}}$ denotes the noise. 
\vera{For single-coil MRI, or when the coil images are combined as a preprocessing step, we have $N_\mathrm{c}=1$ and $\bm s = \bm 1$ in~\eqref{eq:model}.}

The unknown image $\bm x_l\in\cpx^{N_\mathrm{v}}$ for the $l$th echo is problem-dependent, where
\begin{align*}
x_{lj}=
\begin{cases}
m_j&\text{in field map estimation,}\\
m_{\mathrm{w},j}+m_{\mathrm{f},j}\vera{\displaystyle \sum_{p=1}^P\alpha_p e^{i2\pi\Delta_{\mathrm{f},p}t_l}}&\text{in water-fat imaging,}
\end{cases}
\end{align*}
where $\bm m, \bm m_\mathrm{w}, \bm m_\mathrm{f}\in\cpx^{N_\mathrm{v}}$ are respectively the magnetization, water, and fat components, and $\Delta_{\mathrm{f},p}\in\real$ denotes the (known) \vera{frequency shifts of $P$  fat peaks in the multipeak fat model~\cite{reeder2007least},\cite{bydder2008relaxation},\cite{hernando2010robust} with relative amplitudes $\displaystyle \sum_{p=1}^P\alpha_p=1$ that can be estimated and averaged over all fat pixels as a preprocessing step by existing methods~\cite{yu2008multiecho}.}
The goal of the field map estimation problem is to estimate $\bo$ and $\bm x$ given $\bm y$ and $\bm s$.

Assuming the noise $\bm\epsilon$ is zero-mean, white complex Gaussian, the joint maximum-likelihood (ML) estimates of the field map $\bo$ and image $\bm x$ are \vera{the minimizers of the negative log-likelihood as follows:}
$$
\displaystyle \argmin_{\bo,\bm x}\widetilde{\Phi}(\bo,\bm x), \text{ where}
$$
\begin{equation}
\widetilde{\Phi}(\bo,\bm x) = \sum_{j=1}^{N_\mathrm{v}}\sum_{l=1}^L\sum_{c=1}^{N_\mathrm{c}}|y_{clj} - e^{i\omega_j t_l} s_{cj} x_{lj} |^2\mskip2mu.
\label{eq:cost2}
\end{equation}
For a given field map $\bo$, the ML estimate of $\bm x$ has a closed-form expression~\cite{hernando2008joint},\cite{allison2012accelerated} 
that one can substitute into \eqref{eq:cost2} to give a cost function in terms of $\bo$:
\begin{equation}
\displaystyle \Phi(\bo) = \min_{\bm x}\widetilde{\Phi}(\bo,\bm x) 
= \sum_{j=1}^{N_\mathrm{v}}\sum_{m,n=1}^L\sum_{c,d=1}^{N_\mathrm{c}}\phi_{cdmnj}(\omega_j)\mskip2mu,
\label{eq:cost1}
\end{equation}
where
\begin{equation*}
\phi_{cdmnj}(\omega_j)\coloneqq|r_{cdmnj}|\big[1-\cos\big(\angle r_{cdmnj}+\omega_j(t_m-t_n)\big)\big]\mskip2mu,
\end{equation*}
\begin{equation}
r_{cdmnj} \coloneqq \frac{\W_{mn}}{\sum_{c'=1}^{N_\mathrm{c}}|s_{c'j}|^2}s_{cj}s_{dj}^*y_{cmj}^*y_{dnj}\mskip2mu,
\label{eq: r}
\end{equation}
\begin{equation*}
\bm \W\coloneqq\bm\gamma(\bm\gamma^*\bm\gamma)^{-1}\bm\gamma^*\mskip2mu,
\end{equation*}
where $\cdot^*$ denotes the complex conjugate, and $L\times L$ matrix $\bm \W$ is defined in terms of 
\begin{align}
\bm\gamma = 
\begin{cases}
\bm 1&\text{in field map estimation,}\\
\Bigg[\bm 1 \mskip10mu \vera{\displaystyle \sum_{p=1}^P\alpha_p e^{i2\pi\Delta_{\mathrm{f},p}\bm t}}\Bigg]&\text{in water-fat imaging,}
\end{cases}
\label{eq:gamma}
\end{align}
in which $\bm 1$ denotes an all one vector of length $L$, and the exponential is applied element-wise.
In the field map estimation case, this simplifies to $\W_{mn} = 1/L$ $\forall$ $m,n$. 

As $B_0$ field maps tend to be spatially smooth in MRI, we add a regularization term to \eqref{eq:cost1} to form a penalized-likelihood (PL) cost function
\begin{equation}
\displaystyle \Psi(\bo) = \Phi(\bo)+\frac{\beta}{2}\|\bm C\bo\|_2^2\mskip2mu,
\label{eq:cost}
\end{equation}
where $\bm C$ is a first or second order finite difference operator 
\veraa{with optional spatial weights as in~\cite{hernando2010robust}}.
Such regularization has been used in many other prior works~\cite{funai2008regularized},\cite{allison2012accelerated},\cite{ongie2017efficient}.

\section{Efficient Algorithm}\label{sec:alg}
Several approaches have been proposed to solve the field map estimation problem in the single-coil setting, but are demanding in computation or memory.
In particular, a quadratic majorizer with a diagonal Hessian~\cite{funai2008regularized} takes many iterations to converge even for 2D images, and a quadratic majorizer with an optimal curvature that inverts a $N_\mathrm{v} \times N_\mathrm{v}$ Hessian matrix~\cite{allison2012accelerated} is memory-limited to small-scale data.
\veraa{In water-fat imaging,~\cite{hernando2010robust},\cite{sharma2015improving} process data in a single-coil manner using the graph cut method.
Since graph cut requires discretization, \cite{hernando2010robust} proposes to overcome this limitation by additionally running a descent algorithm such as in~\cite{huh2008water}, which considers a quadratic majorizer with a diagonal Hessian that convergences slowly.}

Here, we optimize \eqref{eq:cost} using NCG with a monotonic line search \cite{ongie2017efficient}, and consider a preconditioner with efficient computation and memory storage. 
Our field map estimation procedure is tabulated in the \textbf{Algorithm} below.
For NCG, we choose the Polak-Ribiere update to compute a $\mu^i$ that satisfies the conjugacy condition \cite{polak1969note}.

After estimating the field map $\widehat{\bo}$, we estimate the water and fat components for each voxel in water-fat imaging by applying the closed-form expression~\cite{hernando2008joint} using $\widehat{\bo}$: 
\begin{equation}
\begin{bmatrix}
m_{\mathrm{w},j}\\
m_{\mathrm{f},j}
\end{bmatrix}
= \Big((\bm\gamma\cdot\text{diag}(e^{i\omega_j\bm t})\big)\otimes\bm s_j\Big)^\dagger \bm y_j\mskip2mu,
\label{eq: wf}
\end{equation}
where $\otimes$ denotes the Kronecker product, $(\cdot)^\dagger$ denotes the pseudo inverse, and $\bm s_j\in\cpx^{N_\mathrm{c}}$ denotes the coil sensitivity map for the $j$th voxel.

Next we present our initialization, choice of preconditioner, and derive our iterative monotone line search algorithm in the multi-coil setting. 

\subsection{Initialization}

\begin{figure*}[t!]
\centering
\includegraphics[width=1\textwidth]{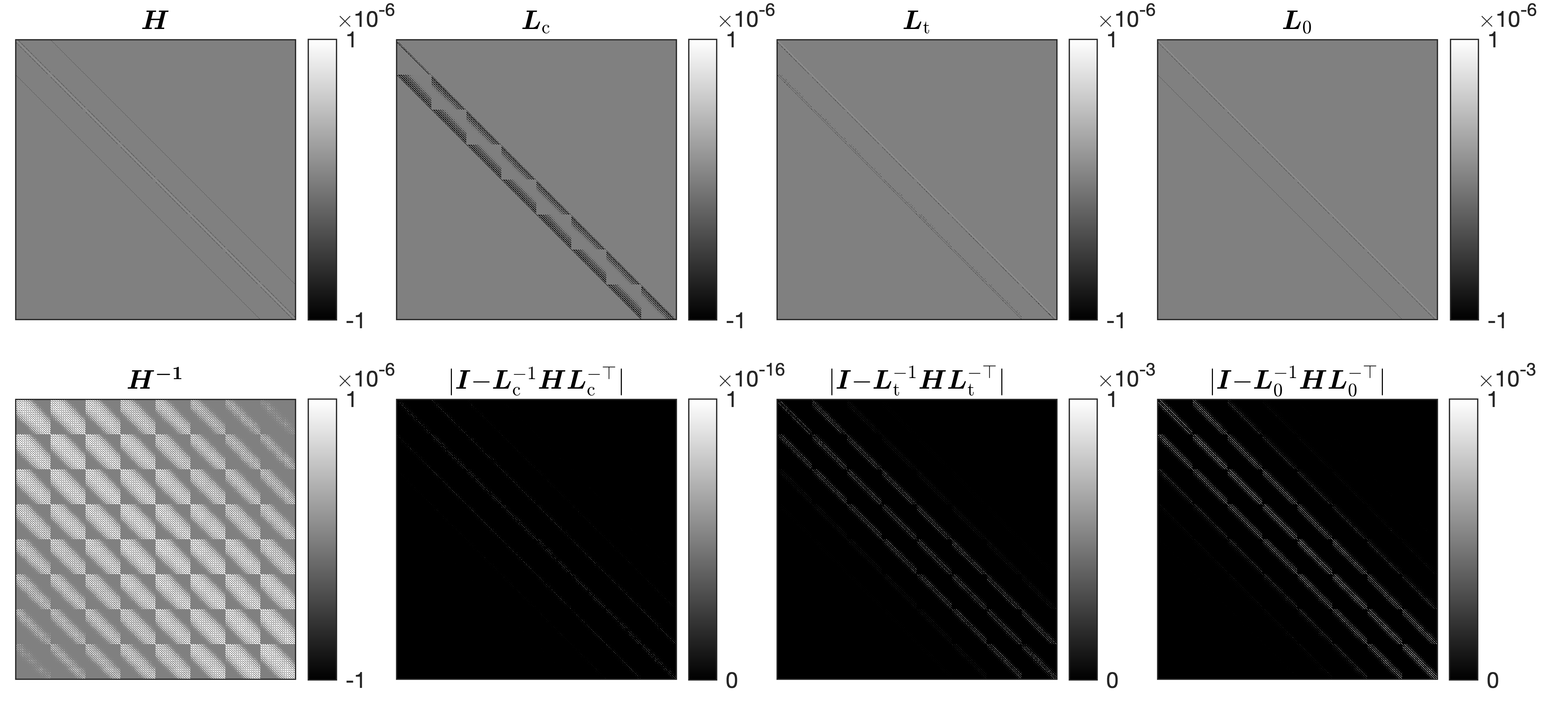}
\caption{Matrix structure of each factorization and the error of its inverse, in a toy problem of size $20\times 16\times 8$.}
\label{fig:ichol}
\end{figure*} 

For field map estimation, we initialize $\bo$ by \vera{a field map computed from the phase of the first two echoes of the coil combined images:}
\begin{equation}
\displaystyle(\omega_j)^0 = \angle \bigg[\bigg(\sum_{c=1}^{N_\mathrm{c}}s_{cj}^*y_{c1j}\bigg)^*\bigg(\sum_{d=1}^{N_\mathrm{c}}s_{dj}^*y_{d2j}\bigg)\bigg]\Big/(t_2-t_1)\mskip2mu.
\label{eq: w0}
\end{equation}

To initialize $\bo$ for water-fat imaging, we follow~\cite{huh2008water} and sweep through a range of \vera{$100$} values from $-|\Delta_\mathrm{f}/2|$ to $|\Delta_\mathrm{f}/2|$ for each voxel, and choose the value with minimal cost \eqref{eq:cost1}, denoted as $\widetilde{\bo}^0$. 
We then run a few CG iterations to minimize a penalized weighted least squares (PWLS) problem
\begin{equation}
\displaystyle\bo^0 = \argmin_{\bo} \sum_{j=1}^{N_\mathrm{v}}\rho_j(\omega_j-\widetilde{\omega}_j^0)^2 + \frac{\beta}{2}\|\bm C\bo\|_2^2\mskip2mu,
\label{eq: w0_wf}
\end{equation}
where the spatial weights
$$\rho_j = \displaystyle\sum_{m,n=1}^L\sum_{c,d=1}^{N_\mathrm{c}}|r_{cdmnj}|$$ 
are given by \eqref{eq: r}.
We then use $\bo^0 $ as our initial estimate in the water-fat case.

\verac{To reduce ambiguity of water and fat assignment, 
one can also consider robust initialization schemes
such as demodulation~\cite{diefenbach2018improving} or magnetization transfer~\cite{samsonov2019resolving}. 
}

\subsection{Preconditioning matrices}\label{subsec:precon}

To accelerate the NCG-based algorithm, given gradient $\bm g^i$ of the cost at the $i$th NCG iteration, we explore a preconditioner $\bm P^i$ with memory efficient implementation of $(\bm P^i)^{-1}\bm g^i$ using an incomplete Cholesky factorization \cite{manteuffel1980incomplete}.
In particular, the gradient $\bm g\in\real^{N_\mathrm{v}}$ is given by
\begin{equation}
\bm g = \nabla \Psi(\bo) = \nabla \Phi(\bo) + \beta \bm C^\top\bm C\bo\mskip2mu,
\label{eq:ls_grad}
\end{equation}
where
\begin{equation*}
\begin{split}
\big(\nabla \Phi(\bo)\big)_j =\sum_{m,n=1}^L\sum_{c,d=1}^{N_\mathrm{c}}
|r_{cdmnj}|(t_m-t_n)^2\\
\cdot\sin\big(\angle r_{cdmnj}+\omega(t_m-t_n)\big)\mskip2mu.
\end{split}
\end{equation*}
The Hessian of the cost \eqref{eq:cost} at the $i$th iteration is the sum of a diagonal matrix and an (approximately, due to the support mask) block Toeplitz with Toeplitz block (BTTB) matrix:
\begin{equation}
\bm H^i = \bm D^i + \beta \bm C^\top\bm C\in\real^{N_\mathrm{v}\times N_\mathrm{v}}\mskip2mu,
\label{eq: precon_hess}
\end{equation}
where $\bm C$ is the finite difference operation and $\bm D^i = \text{diag}(d_j^i)\succeq 0$, where the Hessian of the negative log-likelihood has diagonal elements given by
\begin{equation}
d_j^i = \sum_{m,n=1}^L\sum_{c,d=1}^{N_\mathrm{c}}\kappa_{cdmnj}\big(u_{cdmnj}(\omega_j^i)\big)\mskip2mu,
\label{eq:ls_huber}
\end{equation}
with
\begin{equation*}
\kappa_{cdmnj}(u) = |r_{cdmnj}|(t_m-t_n)^2\frac{\sin(u)}{u}\mskip2mu,\text{ and}
\end{equation*}
\begin{equation}
u_{cdmnj}(\omega) = \big(\angle r_{cdmnj}+\omega(t_m-t_n)\big)\mskip1mu\text{mod}\mskip1mu\pi\mskip2mu.
\label{eq:inner_funs}
\end{equation}

Since the terms $r_{cdmnj}$ and $t_m-t_n$ are shared across iterations, we precompute them at the initialization stage to efficiently calculate the gradient and Hessian at each iteration~$i$.
Note also that $\bm H^i$ is positive definite as long as at least one value of $d_j^i$ is positive (which is true for any nontrivial problem).

Although $\bm H^i$ is sparse and banded, its inverse is approximately full, so directly computing the inverse would require far too much memory. 
To reduce memory, we propose to use a preconditioner that approximates the symmetric Hessian with a LU factorization of the form
\begin{equation}
\bm P^i = \bm L^i(\bm L^i)^{\top}\approx\bm H^i\mskip2mu,
\label{eq: precon}
\end{equation}
where $\bm L^i\in\real^{N_\mathrm{v}\times N_\mathrm{v}}$ is sparse lower triangular, enabling efficient computation (via back-substitution) of $(\bm P^i)^{-1}\bm g^i$ in the precondition step.
Taking advantage of the sparsity and positive definiteness of our Hessian \eqref{eq: precon_hess}, preconditioning with an incomplete Cholesky factorization reduces both computation and memory. 
A popular form of the incomplete Cholesky factorization matches the matrix $\bm H$ on its nonzero set, thus is at least as sparse as $\bm H$.
\vera{Similar preconditioning with incomplete LU factorization has been used for simulating anisotropic diffusion in MRI~\cite{kang2004performance}.}
In practice, for a better approximation one can control the sparsity of the factors by defining a tolerance on the magnitude of the elements of $\bm H$ \vera{(below which entries in the factors are set to zero)}, with the trade-off between approximation accuracy and memory storage.

\begin{table}[t!]
\centering
\begin{tabular}{ c|c|c|c|c|c } 
   & $\bm H$ & $\bm{H}^{-1}$ & $\bm{L}_\mathrm{c}$ & $\bm{L}_\mathrm{t}$ & $\bm{L}_\mathrm{0}$  \\ 
  \hline
Number of nonzeros ($\times 10^5$) & 1.67 & 655 & 72.5 & 1.77 & 0.96  \\
\hline
Storage (megabytes) & 0.31 & 100.1 & 11.9 & 0.53 & 0.27\\
\hline
NRMSE &  &  & 3e-16 & 4e-3 & 3e-2
\end{tabular}

\vspace*{.1in}

\caption{Number of nonzero elements, memory usage, and NRMSE of the inverse of each factorization in a toy problem of size $20\times 16\times 8$.}
\label{tab: ichol}
\end{table}

\begin{figure}[t!]

\vspace*{-.2in}

\centering
\includegraphics[width=.38\textwidth]{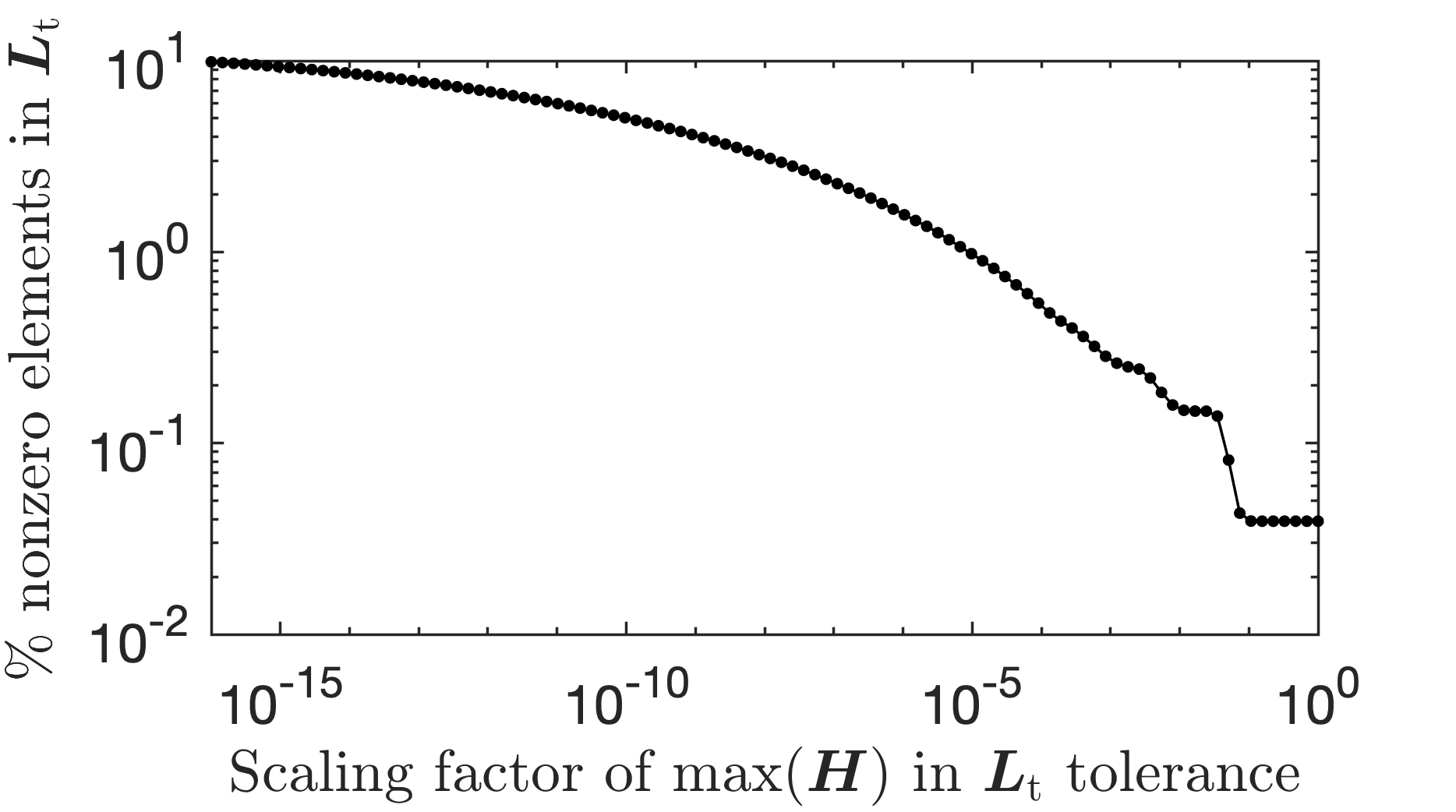}
\caption{\vera{Change of sparsity of $\bm{L}_\mathrm{t}$ with respect to the scaling factor of $H_\mathrm{max}$ in its tolerance.}}
\label{fig:ichol_sparsity}
\end{figure}

Fig.~\ref{fig:ichol} illustrates the memory improvement by a toy problem of image size $20\times 16\times 8$, where we compute $\bm H = \bm D + \beta \bm C^\top\bm C$ and its inverse, with randomly chosen diagonal elements $d_j\in (0,0.1)$ and $\beta = 0.1$.
Fig.~\ref{fig:ichol} considers the incomplete Cholesky factorization without tolerance, denoted $\bm{L}_\mathrm{0}$, and with a tolerance of $H_\mathrm{max} \times 10^{-3}$, denoted $\bm{L}_\mathrm{t}$, where $H_\mathrm{max}$ is the element in $\bm H$ with maximum magnitude.
Fig.~\ref{fig:ichol} shows the sparse structure of $\bm H$, its nonsparse inverse $\bm H^{-1}$, and the Cholesky factorizations as well as their approximation errors.
Table~\ref{tab: ichol} shows the number of nonzero elements of each matrix, their memory storage, and their errors that affect the convergence rate, using the normalized root mean square error (NRMSE) $\|\bm I - \bm{L}^{-1} \bm H \bm{L}^{-\top} \|_\mathrm{F} /\sqrt{N_\mathrm{v}}$ for each factorization $\bm L$ in our example.
\vera{Fig.~\ref{fig:ichol_sparsity} illustrates how the sparsity of $\bm{L}_\mathrm{t}$ changes with respect to its tolerance by showing the percentage of nonzero elements in $\bm{L}_\mathrm{t}$ versus the scaling factor of $H_\mathrm{max}$ in the tolerance.}

For memory storage in this case, the number of nonzero elements in the incomplete Cholesky factor without tolerance $\bm{L}_\mathrm{0}$ is more than $70$ times less than that in the (complete) Cholesky factor $\bm{L}_\mathrm{c}$, with more than $40$ times memory saving.
In general, we observe (by the banded structures) that the number of nonzero elements of $\bm{L}_\mathrm{c}$ is lower bounded by $(N_\mathrm{v}-N_\mathrm{x}N_\mathrm{y})*N_\mathrm{x}N_\mathrm{y}$, while that of $\bm{L}_\mathrm{0}$ is upper bounded by $4N_\mathrm{v}$. 
This leads to the generalization that $\bm{L}_\mathrm{0}$ is at least $(N_\mathrm{v}-N_\mathrm{x}N_\mathrm{y})/(4N_\mathrm{z})$ times more sparse than $\bm{L}_\mathrm{c}$, which scales significantly with the problem size.
The storage of the incomplete Cholesky factor with tolerance $\bm{L}_\mathrm{t}$ depends on the tolerance, and with the choice of tolerance here we observe a $40$ times fewer nonzero values, saving memory by a factor of more than $20$
compared with $\bm{L}_\mathrm{c}$.

The trade-off with a sparser factorization, however, is a worse approximation error. 
This is reflected in the error matrices in Fig.~\ref{fig:ichol} and the NRMSE in Table~\ref{tab: ichol}.
While $\bm{L}_\mathrm{0}$ has lower memory usage than $\bm{L}_\mathrm{t}$, the inverse is a \vera{worse} approximation to $\bm H^{-1}$.
In practice, nevertheless, both incomplete factorizations $\bm L\bm L^\top$ are positive definite, so as preconditioners they provide \vera{a descent direction} in addition to storage advantage, whereas storing $\bm{L}_\mathrm{c}$ is infeasible for realistically sized 3D datasets.

\begin{table}[b!] 
\normalsize
\begin{tabular}{ l } 
\hline
\textbf{Algorithm: }Preconditioned NCG-MLS\\
\hline
\textbf{Inputs: } \\ 
\tab $\bm y$, $\bm s$, $\bm t$, $\bm C$, $\beta$\\
\textbf{Intialization: } \\ 
\tab $\displaystyle\bo^0$ by \eqref{eq: w0} or \eqref{eq: w0_wf}\\
\tab $\bm z^0 = -\nabla \Psi(\bo^0)$\\
\tab $\alpha^{(0)} = 0$\\
\tab precompute $r_{cdmnj}$ by~\eqref{eq: r} and $t_m-t_n$\\ 
\textbf{for} $ i = 0, 1, \ldots, N-1$ \textbf{do}\\
\tab compute gradient $\bm g^i = \nabla \Psi(\bo^i)$ with \eqref{eq:ls_grad}\\
\tab precondition $\bm p^i = ( \bm P^i)^{-1}\bm g^i$ with \eqref{eq: precon}\\
\tab compute $\mu^i$ with conjugacy \\
\tab search direction $\bm z^{i+1} = \bm p^i + \mu^i\bm z^i \in\real^{N_\mathrm{v}}$\\
\tab \textbf{for} $ k = 0, 1, \ldots, N_i-1$ \textbf{do}\\
\tab\tab update step size $\alpha^{(k+1)}$ by \eqref{eq:ls_update}\\
\tab \textbf{end for}\\
\tab update $\bo^{i+1} = \bo^i + \alpha^{(N_i)}\bm z^{i+1}$\\
\textbf{end for}\\
\textbf{output:} $\bo^N$\\
\hline
\end{tabular}
\end{table}

\subsection{Monotonic step size line search}\label{subsec:mono}
With a search direction given by NCG, the choice of step size is important for convergence of the algorithm.
To avoid multiple function evaluations required by backtracking line search algorithms \cite{nocedal2006numerical}, we implement a recursive line search algorithm using a quadratic majorizer with an optimal curvature, which guarantees monotone decrease of the cost function \cite{fessler1999conjugate}.

In the line search step, given a current field map estimate $\bo^i$ and a search direction $\bm z^i \in\real^{N_\mathrm{v}}$, we aim to find a step size that minimizes the cost \eqref{eq:cost}:
\begin{equation*}
\widehat{\alpha}= \displaystyle\argmin_\alpha f(\alpha)\mskip2mu,\text{ where}
\end{equation*}
\begin{equation}
f(\alpha) = \Phi(\bo^i  + \alpha \bm z^i )\\
+\frac{\beta}{2}\|\bm C (\bo^i  + \alpha \bm z^i ) \|_2^2\mskip2mu,
\label{eq:ls_cost}
\end{equation}

We iteratively minimize the nonconvex problem \eqref{eq:ls_cost} using a quadratic majorizer based on Huber's method \cite[p.~184]{huber1981robust} at the $k$th inner iteration (dropping outer iteration $i$ for brevity):\\
\begin{align*}
q_k(\alpha) &= \Phi(\bo + \alpha^{(k)} \bm z)\\
&+\bm z^\top\nabla \Phi(\bo+ \alpha^{(k)} \bm z)(\alpha-\alpha^{(k)} )\\
&+\frac{1}{2}d^{(k)}(\alpha-\alpha^{(k)} )^2
+\frac{\beta}{2}\|\bm C (\bo + \alpha \bm z) \|_2^2\mskip2mu,
\end{align*}
where the optimal curvature is given by~\cite{allison2012accelerated}
\begin{equation*}
d^{(k)} =\sum_{j=1}^{N_\mathrm{v}}|z_j|^2 d_j^{(k)}\mskip2mu,\text{where}
\end{equation*}
\begin{equation}
d_j^{(k)} = \sum_{m,n=1}^L\sum_{c,d=1}^{N_\mathrm{c}}\kappa_{cdmnj}\big(u_{cdmnj}(\omega_j+\alpha^{(k)}z_j)\big)\mskip2mu,
\label{eq:ls_huber2}
\end{equation}
with
$\kappa_{cdmnj}(\cdot)$ and $u_{cdmnj}(\cdot)$ defined in \eqref{eq:inner_funs}.

Using one step of Newton's method on the quadratic majorizer $q_k(\alpha)$ gives the step size update
\begin{align}
\alpha^{(k+1)}& = \alpha^{(k)} - \frac{\frac{\partial}{\partial \alpha} q_k( \alpha^{(k)})}{ \frac{\partial^2}{\partial \alpha^2} q_k( \alpha^{(k)})} \nonumber\\
&= \alpha^{(k)} -\frac{\frac{\partial}{\partial \alpha}f(\alpha^{(k)})}{d^{(k)}+\beta \|\bm C\bm z\|^2_2} \mskip2mu.
\label{eq:ls_update}
\end{align}

We implement \eqref{eq:ls_update} efficiently by computing $ \|\bm C\bm z\|^2_2$ only once per outer NCG iteration $i$.
Since the majorizer satisfies $q_k(\alpha)\geq f(\alpha)$ for all step size $\alpha$ and inner line search iteration $k$, the update \eqref{eq:ls_update} guarantees monotonic decrease of the cost \eqref{eq:ls_cost}.

\section{Results}\label{sec:res}
We investigated our algorithm and its efficiency with two multi-echo field map estimation experiments and two water-fat imaging experiments.
Due to the large data size, memory intensive methods with a direct solver using the full Hessian are excluded from our experiments.
In particular, we compare our incomplete Cholesky preconditioner (NCG-MLS-IC) method versus a quadratic majorizer update with diagonal Hessian (QM) \cite{funai2008regularized} and versus the NCG algorithm without any preconditioner (NCG-MLS) and with a diagonal preconditioner (NCG-MLS-D) \cite{allison2012accelerated}.
\vera{In addition, we used the Poblano toolbox~\cite{poblano} to compare the convergence of the quasi-Newton (QN) and truncated Newton (TN) methods in our simulations.}

For each dataset, we define a mask using the convex hull of all voxels \vera{that contribute to the signal} (with coil-combined image magnitude thresholded below by $0.1 y_\text{max}$, where $y_\text{max}$ denotes the maximum image magnitude in the coil-combined image for the first echo time.), with a dilation of two voxels.
We then computed $\bo$ within the mask, and tuned the regularization parameter $\beta$ by sweeping across a range of values.
All our experiments used MATLAB R2020a, with a 2.4-GHz dual-core Intel Core i7.
The MATLAB code that reproduces the experiments with our efficient algorithm will be available as part of the Michigan Image Reconstruction Toolbox (MIRT) \cite{mirt}.

\subsection{Brain Simulation}\label{sec:res_simu}

We first simulated a 3D brain dataset with 40 $64\times 64$ slices, 4 simulated coils and 3 echo times $t_l = 0,2,10$ ms, with added complex Gaussian noise so that the SNR $\approx 20$ dB.
To generate multi-coil data, we simulated coil sensitivity maps with 4 coils based on \cite{ grivich2000magnetic} using the MIRT.
We set $\beta = 2^{-4}$ with first order regularization to achieve visual resemblance to the ground truth field map.
In light of the trade-off between storage and approximation error discussed in Section~\ref{subsec:precon}, we explored preconditioners using the incomplete Cholesky factorization both without tolerance (NCG-MLS-IC-0) and with a tolerance of  $H^i_\mathrm{max} \times 10^{-3}$ for each iteration $i$ (NCG-MLS-IC).

\begin{figure}[b!]
\centering
\vspace{-.2in}
\includegraphics[width=.45\textwidth]{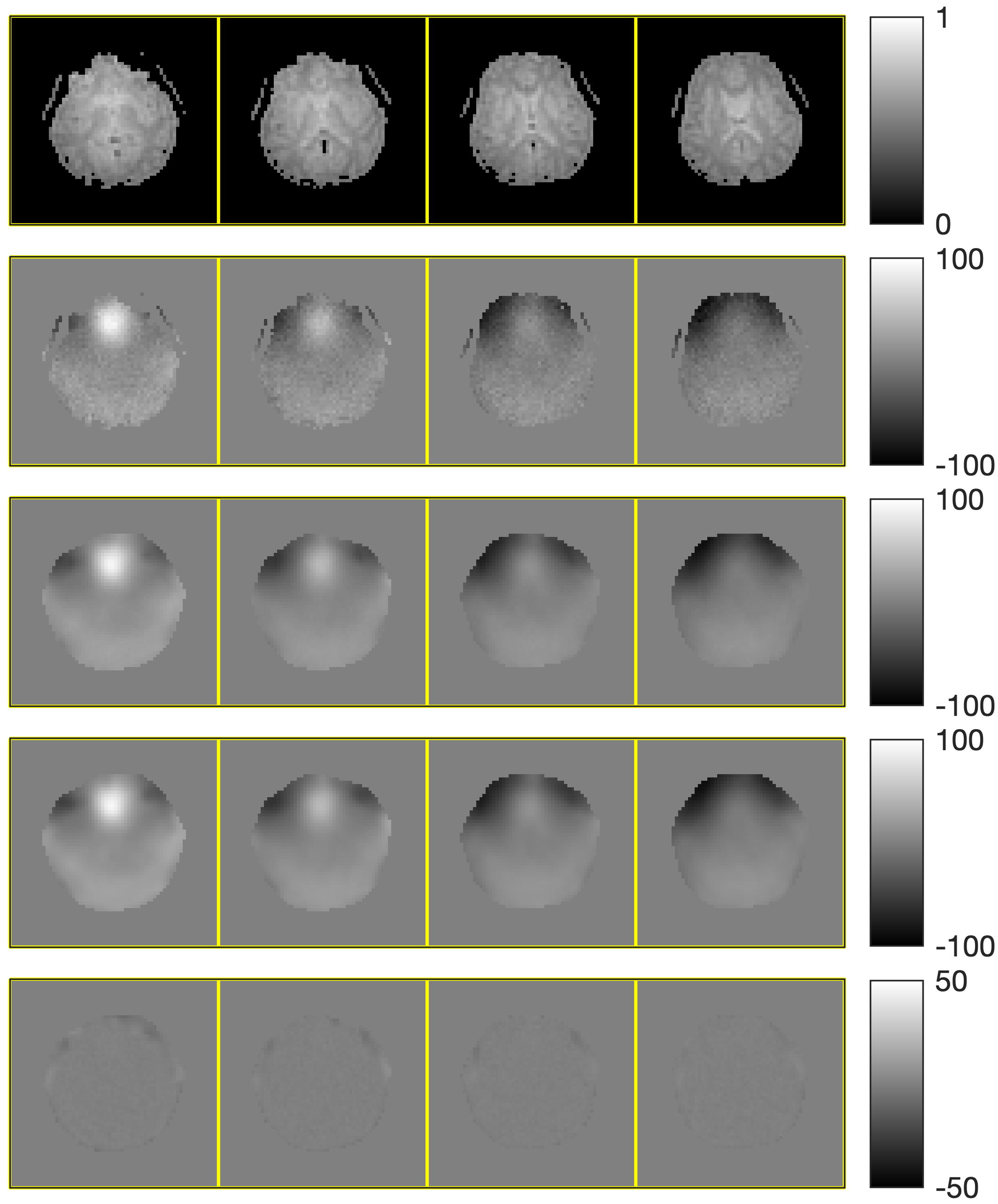}
\caption{Top to bottom: selected slices of coil-combined simulation image, initial field map (in Hz), regularized field map estimate $\widehat{\bo}$, ground truth field map $\bo_\text{true}$, and error $|\widehat{\bo}-\bo_\text{true}|$.}
\label{fig:simu_ims}

\includegraphics[width=.48\textwidth]{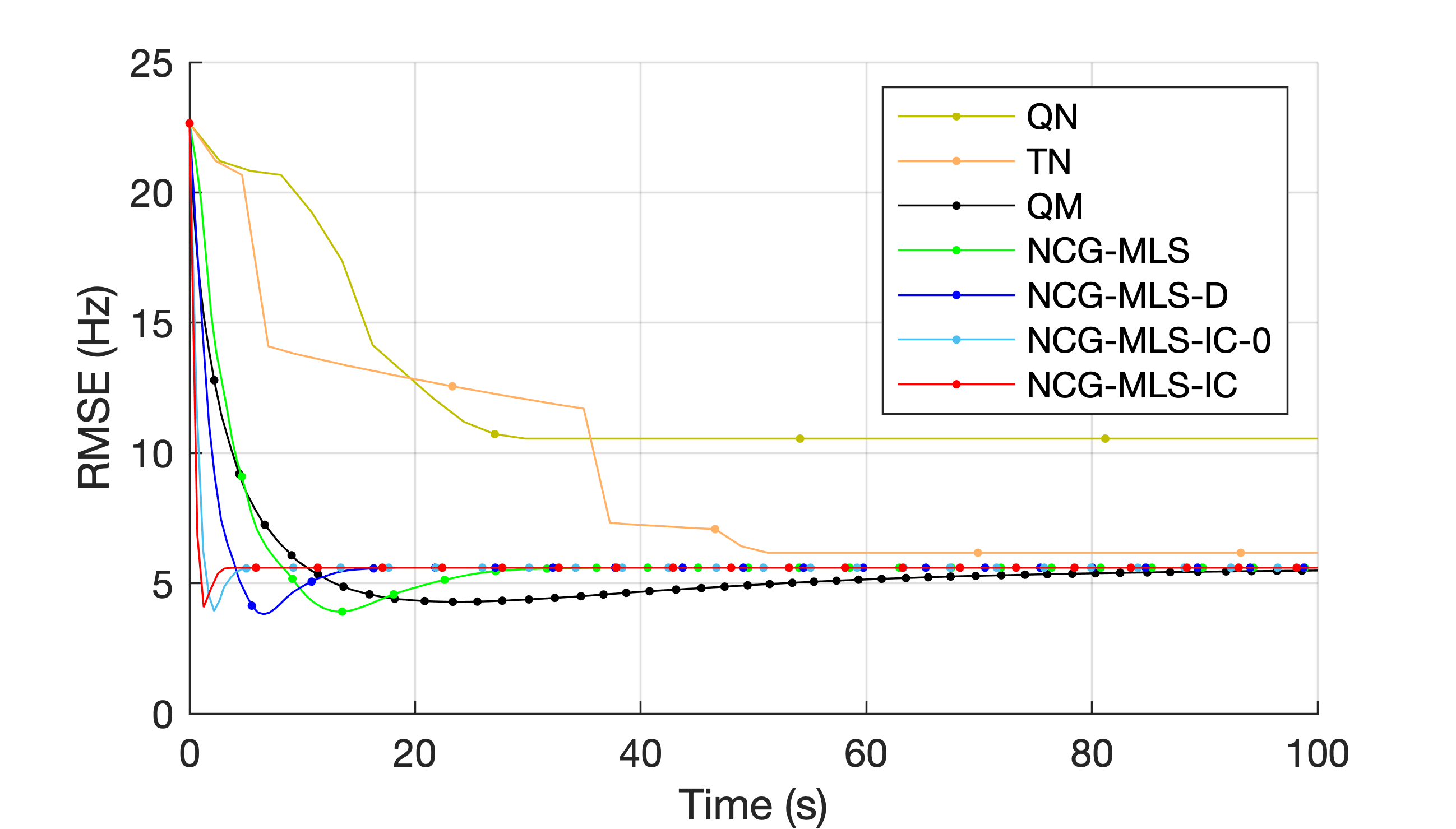}
\caption{\vera{RMSE versus wall time of seven algorithms used in simulation. Every 10 iteration is marked by a dot.}}
\label{fig:simu_nrmsd}
\end{figure}

Fig.~\ref{fig:simu_ims} shows four selected slices, their initial field map, and the regularized estimate by our algorithm.
To examine the speed of convergence, we plot the \vera{root mean square error (RMSE) $\|\bo^i-\bo_\text{true}\|_2/\sqrt{N_\mathrm{v}}$ versus wall time in Fig.~\ref{fig:simu_nrmsd}. 
The RMSE plots show that the QM and all the NCG-MLS methods converge to RMSE $\approx 5.6$ Hz, though going through a slightly lower RMSE in the iterative process. 
Both the quasi-Newton and the truncated Newton methods converge to minimizers with higher RMSE, hence we omitted their comparison in the phantom experiment below. 
The plots show a significant computational gain of NCG-MLS preconditioned with the incomplete Cholesky factorization over all the other methods.
}
We also observe that using a nonzero tolerance in the incomplete Cholesky factorization gives a slightly faster convergence than not using one, hence we adopt that choice for the NCG-MLS-IC implementations in our next experiments.

\subsection{Phantom Dataset}\label{sec:res_phantom}

\begin{figure}[b!]
\centering
\includegraphics[width=.45\textwidth]{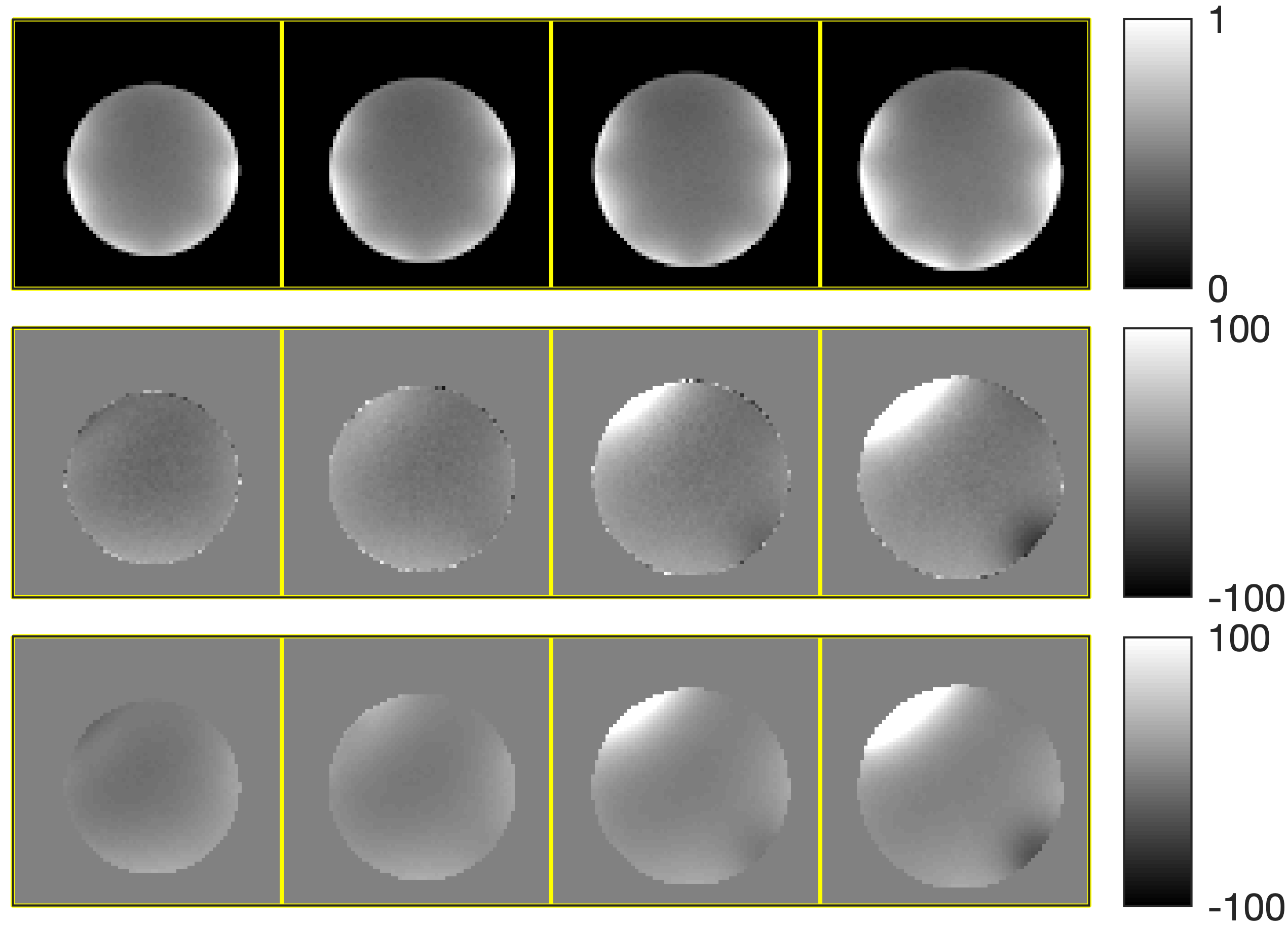}
\caption{Top to bottom: selected slices of coil-combined phantom image, initial field map (in Hz), and regularized field map estimate.}
\label{fig:phantom_ims}

\includegraphics[width=.48\textwidth]{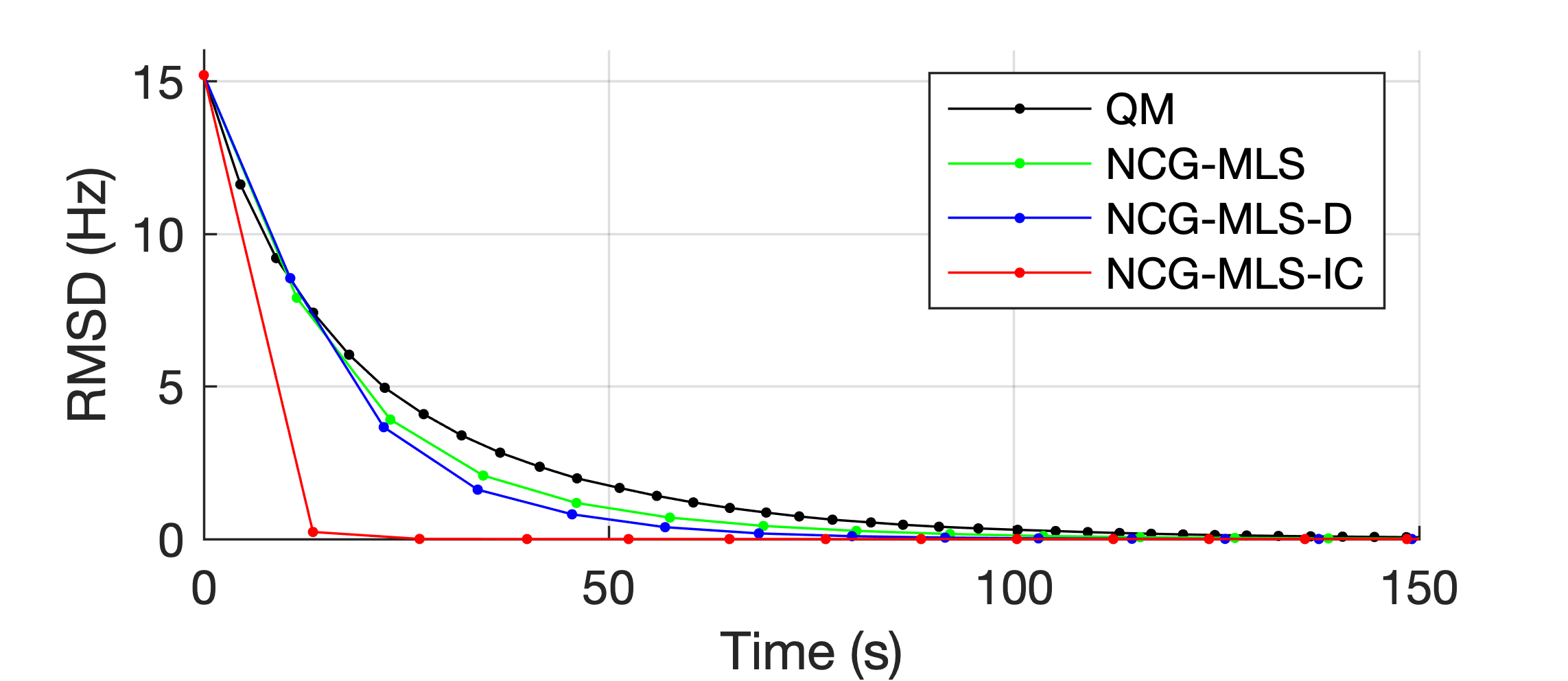}
\caption{RMSD versus wall time of four algorithms used in the phantom experiment. Every iteration is marked by a dot.}
\label{fig:phantom_nrmsd}
\end{figure} 

\begin{table}[b!]
\centering
\begin{tabular}{ c|c|c|c|c} 
   & QM & NCG-MLS & NCG-MLS-D & NCG-MLS-IC  \\ 
    \hline
Time (s) & 96 & 81 & 69 & 4.5\\
 \hline
vs. IC time & 21$\times$ & 18$\times$ & 15$\times$ & 
\end{tabular}

\vspace*{.1in}

\caption{Time for each method to reach an RMSD below $0.5$ Hz, and their relative proportions to the time taken by NCG-MLS-IC.}
\label{tab: phantom}
\end{table}

Our second experiment uses a Function Biomedical Informatics Research Network (FBIRN) phantom \cite{keator2016function} with two pieces of metal staple to induce field inhomogeneity, collected on a GE MR750 3T scanner with a 32-channel Nova Head Coil receiver. 
This dataset has size $74\times 74\times 10$ with $3$~mm$^3$ isotropic voxel size, TR = $10.5$~ms, with 3 echo times $t_l = 0,1,2.3$~ms.
We computed coil sensitivity maps using ESPIRiT~\cite{uecker2014espirit}, and set $\beta = 2^{-3}$ with first-order finite difference regularizaiton.

Fig.~\ref{fig:phantom_ims} shows four selected slices, their initial field map, and the regularized estimate by our algorithm.
\veraa{To compare convergence, we computed the root mean square difference (RMSD) $\|\bo^i-\bo^\infty\|_2/\sqrt{N_\mathrm{v}}$ to the converged
$\bo^{\infty}$ of the QM method.}
The RMSD plots in Fig.~\ref{fig:phantom_nrmsd} show that our algorithm  converges much faster than the other three, reaching $0.33$~Hz RMSD in $1$ iteration, and $0.005$~Hz RMSD in $2$ iterations.
Since this 3D dataset has a more realistic problem size than the simulated data, we quantify the convergence speedup by comparing the time it takes for each method to reach an RMSD below $0.5$~Hz.
Table~\ref{tab: phantom} shows that our NCG-MLS algorithm with an incomplete Cholesky preconditioner provides \vera{a speedup of 15 times} from NCG-MLS with a diagonal preconditioner, 18 times from that without a preconditioner, and 21 times from the quadratic majorizer implementation.

\subsection{\vera{Cardiac Water-Fat Simulation}}\label{sec:2d_wf_simu}

\begin{figure}[b!]
\centering
\includegraphics[width=.44\textwidth]{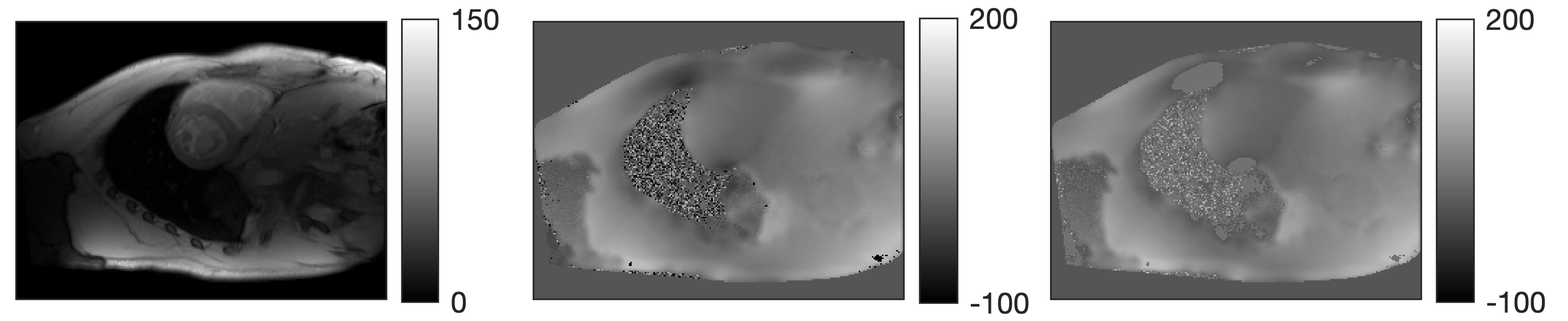}
\caption{\vera{Left to right: simulated image for the $1^\text{st}$ echo, initial field map $\widetilde{\bo}^0$ (in Hz) by voxel-wise estimation, and initial fieldmap $\bo^0$ by PWLS~\eqref{eq: w0_wf}.}}
\label{fig:2d_wf_init}
\vspace*{.1in}
\includegraphics[width=.44\textwidth]{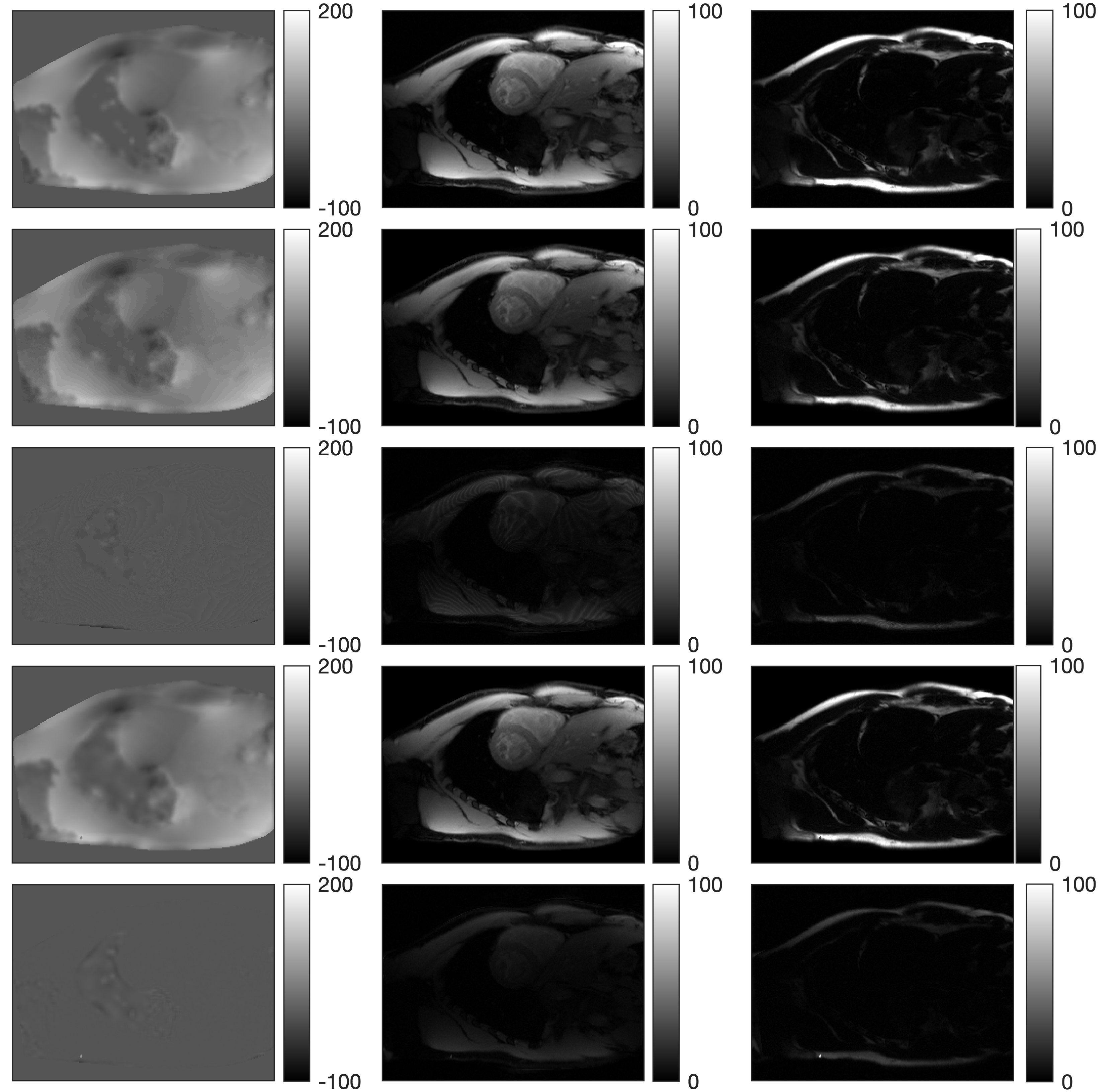}
\caption{\vera{$1^\text{st}$ row: ground truth field map, water, and fat images. 
$2^\text{nd}$ and $3^\text{rd}$ row: graph cut estimates and their error images. 
$4^\text{th}$ and $5^\text{th}$ row:  NCG-MLS-IC estimates and their error images. }}
\label{fig:2d_wf_ims}
\end{figure} 

\begin{figure}[b!]
\centering
\includegraphics[width=.48\textwidth]{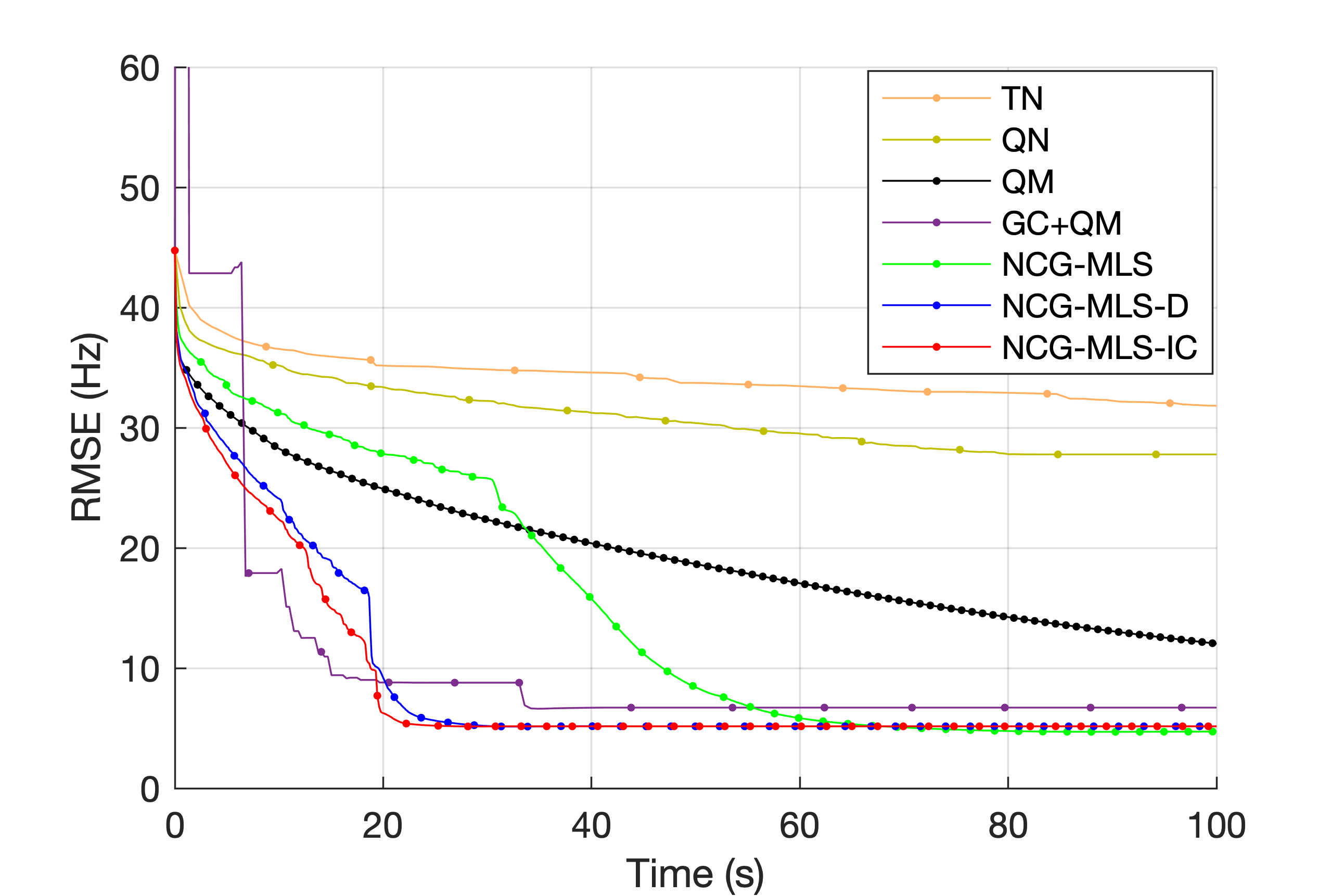}
\caption{\vera{Field map RMSE versus wall time of seven algorithms used in the water-fat simulation. Every 20 iterations is marked by a dot.}}
\label{fig:2d_wf_nrmsd}
\end{figure} 

\vera{For water-fat imaging, we first} \veraa{performed a cardiac simulation based on one of the 8-echo datasets} \vera{used in the ISMRM Fat-Water Toolbox~\cite{fatwater}.
Since} \veraa{implementations in the toolbox work only for 2D datasets, and} \vera{coil combination such as~\cite{walsh2000adaptive} is often used in practice, we illustrate the flexibility of our algorithm in a 2D coil-combined case by simply setting the number of coils $N_\mathrm{c}=1$ and the coil sensitivitiy map  $\bm s = \bm 1$. 
We also consider the multipeak model in water-fat imaging. 
}

\vera{This dataset has size $256\times 192$ with 8 echo times from $1.5$ to $17.4$ ms (each $2.3$ ms apart).
We generated ground truth field map and water and fat images using  golden section search with multiresolution~\cite{lu2008multiresolution}. }
\veraa{We used the same values $\{\alpha_p\}$ and $\{\Delta_{\mathrm{f},p}\}$ as in the toolbox implementations both for simulating images
with 8 echo times using the model~\eqref{eq:model}
and for estimation.
For comparison, we also ran the graph cut (GC) method~\cite{hernando2010robust} using the same cost~\eqref{eq:cost} with second-order finite differences as in \cite{hernando2010robust}, and $\beta = 2^{-7}$ as the regularization parameter.
}

\vera{Fig.~\ref{fig:2d_wf_init} shows the first echo image, the initial field map $\widetilde{\bo}^0$ by voxel-wise estimation, and the initial $\bo^0$ after $10$ CG iterations of PWLS minimization~\eqref{eq: w0_wf}.
Fig.~\ref{fig:2d_wf_ims} shows the ground truth field map, water and fat images, and the estimates and error images by the graph cut and by our algorithm.
Compared with graph cut, our algorithm achieves slightly lower NRMSE on the water image ($20.09\%$ vs. $23.57\%$) and the fat image ($20.93\%$ vs. $23.43\%$), with lower final RMSE on the field map, shown in Fig.~\ref{fig:2d_wf_nrmsd}. 
To explore a combination suggested by~\cite{hernando2010robust}, we ran 100 graph cut iterations followed by 100 optimal transfer iterations using a quadratic majorizer~\cite{huh2008water}.
\verab{We used the implementation in the toolbox~\cite{fatwater} which did not precompute $r_{cdmnj}$ by~\eqref{eq: r}.}
Fig.~\ref{fig:2d_wf_nrmsd} shows the graph cut RMSE curve jumps up (to $615$ Hz) on its first iteration, and converges to its own minimizer. The subsequent quadratic majorizer update lowers the RMSE further, which opens a promising future direction of \verab{combining graph cut with
the faster NCG-MLS-IC with precomputation of common terms.}
}\vera{
Fig.~\ref{fig:2d_wf_nrmsd} also shows the truncated Newton and quasi-Newton methods again converge to their minimizers with higher RMSE.
We omitted all methods with higher RMSE in the real data experiment below.
}

\subsection{Ankle Water-Fat Dataset}\label{sec:res_wf}

\begin{figure}[t!]
\centering
\includegraphics[width=.44\textwidth]{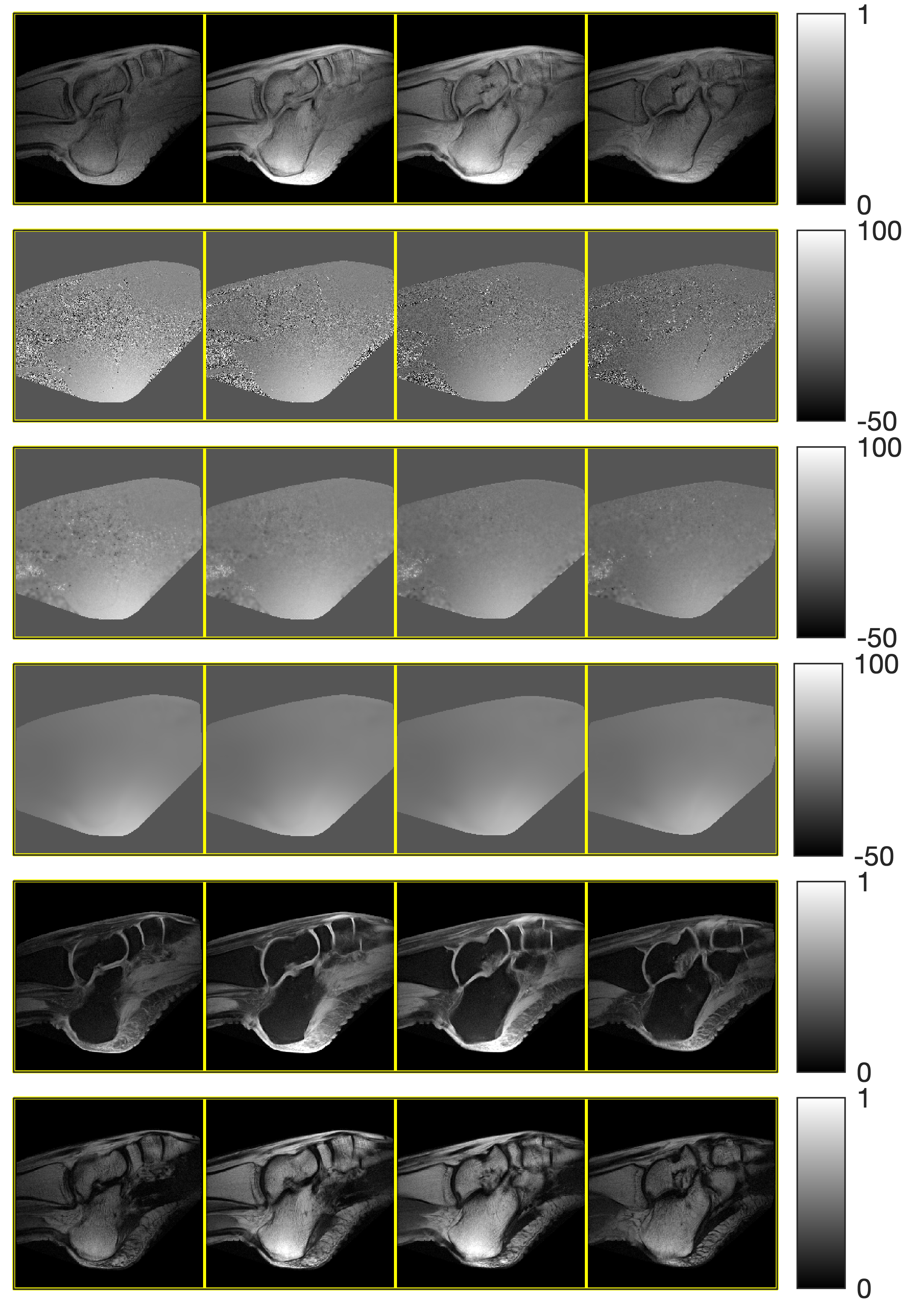}
\caption{Top to bottom: coil-combined water-fat image for the $1^\text{st}$ echo, initial field map $\widetilde{\bo}^0$ (in Hz) by voxel-wise estimation, initial fieldmap $\bo^0$ by PWLS~\eqref{eq: w0_wf}, regularized field map estimate, estimated water image, and estimated fat image.}
\label{fig:invivo_wf_ims}

\includegraphics[width=.48\textwidth]{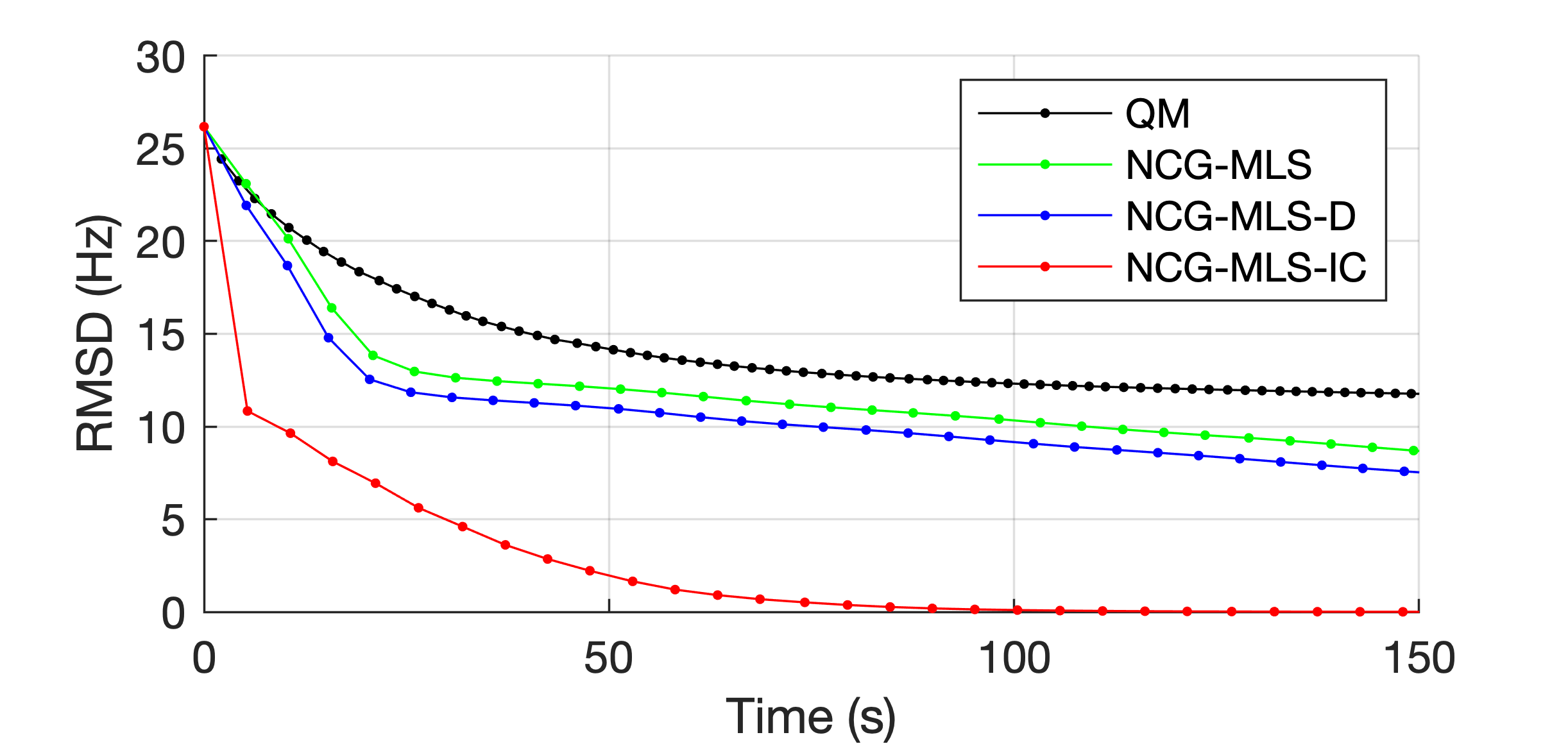}
\caption{Field map RMSD versus wall time of four algorithms used in the water-fat experiment. Every iteration is marked by a dot.}
\label{fig:invivo_wf_nrmsd}
\end{figure} 

\vera{We further illustrate our algorithm in the 3D multi-coil setting using an ankle dataset} from the ISMRM Fat-Water Separation Dataset~\cite{fatwater}.
This dataset has 4 $256\times 256$ slices, 8 coils and 3 echo times $t_l = 2.2, 3, 3.8$ ms, in a 3T scanner that corresponds to a single $\Delta_\mathrm{f}\approx 440$~Hz.
We chose $\beta = 2^{-10}$ with first-order finite difference regularization to achieve visual separation of water and fat components. 

Fig.~\ref{fig:invivo_wf_ims} shows the first echo image, the initial field map $\widetilde{\bo}^0$ by voxel-wise estimation, the initial $\bo^0$ after $10$ CG iterations of PWLS minimization~\eqref{eq: w0_wf}, and the regularized estimate by our algorithm.
For completeness, Fig.~\ref{fig:invivo_wf_ims} also shows the estimated water and fat images using \eqref{eq: wf}, which achieve a visual separation of the two components.
However, it is worth emphasizing that our main interest is in the speed of finding a minimizer of the problem \eqref{eq:cost}. 
\veraa{In this case, since QM converged to a different local minimum than the other three methods, we computed the RMSD to
$\bo^{\infty}$ of the NCG-MLS method (without preconditioner).}
The RMSD plots in Fig.~\ref{fig:invivo_wf_nrmsd} show a significant computational gain of our algorithm over the other algorithms.

\section{Conclusion}\label{sec:con}
This paper presents an efficient algorithm for both multi-echo field map estimation and water-fat imaging problem in the 3D \verac{multi-coil} MRI setting.
Given the nonconvex cost function, our algorithm uses the nonlinear conjugate gradient method with a preconditioner based on an incomplete Cholesky factorization, and a monotonic step size line search based on a quadratic majorizer with optimal curvatures.
This is the first work to use the incomplete Cholesky factorization as a preconditioner for multi-coil field map estimation.
Experiments with simulation and real data show that our method has faster convergence than existing memory-efficient methods. 
 
\section*{Acknowledgement}
We thank Jon-Fredrik Nielsen at the University of Michigan for the multi-echo 3D \verac{multi-coil} MRI phantom data.

\end{document}